\documentclass[doublespacing]{elsart}

\usepackage{amssymb}
\usepackage{epsfig}

\begin{document}

\begin{frontmatter}



\title{A Dual Gate Spin Field Effect Transistor With Very Low
Switching Voltage and Large ON-to-OFF Conductance Ratio}

\author{J. Wan$^a$, M. Cahay$^a$ and S. Bandyopadhyay$^b$}
\address[label1]{Department of Electrical and Computer Engineering, University 
of Cincinnati, Cincinnati, OH 45221, USA}
\address[label2]{Department of Electrical and Computer Engineering, Virginia 
Commonwealth 
University, Richmond, VA 23284, USA}

\begin{abstract}
We propose and analyze a novel dual-gate Spin Field Effect Transistor (SpinFET) 
with  half-metallic ferromagnetic source and drain contacts. 
The transistor has two gate pads that can be biased independently. It can be 
switched ON or OFF with a few mV change in
the differential bias between the two pads, resulting in extremely low dynamic 
power dissipation during switching. The ratio of ON to OFF conductance remains 
fairly large ($\sim$ 60)  
up to a temperature of 10 K. This device also has excellent inverter 
characteristics, making it attractive  for applications in low power and high 
density Boolean logic circuits.
\end{abstract}

 \end{frontmatter}

\newpage
\section{Introduction}
The Spin Field Effect Transistor (SpinFET) proposed by Datta and Das 
\cite{datta} and its various clones (see, for example, \cite{loss,flatte}) all 
operate on the principle of modulating
the transmission of electrons through the device by controlling the spin orbit 
interaction strength in the channel with a gate potential.
Here, we describe a different type of SpinFET where the gate potential modulates 
transmission resonances in the channel, instead of the spin orbit interaction.  
By changing the gate potential, we can move the Fermi level in the device from 
regions of maximum transmission (pass bands) to regions of minimum transmission 
(stop bands), thus modulating the device conductance. An immediate benefit of 
this approach is a much reduced switching voltage. Spin orbit interactions in 
technologically important semiconductors are typically weak and weakly sensitive 
to external potentials \cite{nitta}. As a result, a large swing in the gate 
potential (few Volts) is often required to turn a short channel 
SpinFET of the traditional type on or off \cite{bandy_apl}. That results in 
considerable dynamic energy dissipation during switching. In contrast, our 
strategy is to modulate transmission resonances (Ramsauer and Fano-type 
resonances) that occur in the channel of traditional SpinFETs 
\cite{cahay1,cahay2,cahay3} with a gate potential. Since the resonance widths 
(in energy) are very small, a few mV change in the gate potential can take the 
device from on-resonance to off-resonance and switch it on or off. This approach 
results in a very small switching voltage (a few mV instead of the few Volts 
required in traditional SpinFETs) resulting in much reduced dynamic energy 
dissipation \cite{cahay3}.

The downside of this strategy is that the temperature of operation must be low 
so that the thermal broadening in the electron energy remains much smaller than 
the widths of the transmission resonances. Typically, this  limits the 
temperature of operation to $\sim$ 1 K for realistic device parameters 
\cite{cahay3}. Here, we show that this problem can be somewhat mitigated by 
using a dual-gate configuration (inter-digitated gate) which allows us to apply 
two different gate biases on different regions of the channel. This will allow 
us to engineer the conduction band profile within the channel in a way that will 
broaden the transmission resonances and extend the temperature of operation to  
higher temperatures. Two additional benefits are: (1) the ratio of on-to-off 
conductance remains large over an extended temperature range, and (2) the device 
exhibits sharp turn-on or turn-off characteristics, making it ideal for 
application as an inverter. The inverter type behavior is much sought after in 
Boolean logic circuits.

\section{Theory}

The proposed SpinFET with the dual-gate configuration is shown
in Fig. 1. The structure consists of one or several  quantum wires in parallel  
(each with the lowest subband occupied) sandwiched between two 
half-metallic ferromagnetic contacts. The use of parallel channels increases the 
overall conductance of the
device. 

The potential on the outer
gate ($V_{g1} $) is selected such that underneath this gate, the conduction band 
edge in the channel is at a location
$\Delta{E_c}^{\star}$ above the bottom of the majority spin band in the 
ferromagnets (see Fig. 2). The bias on the inner gate ( $V_{g2} $) is varied to 
modulate the conduction band edge underneath this gate by an amount $\pm \delta 
(\Delta{E_c})$ around the quiescent value. Accordingly, the conduction band 
profile 
$\Delta{E_c} (x)$ along the channel will form the basic unit of a superlattice 
along the direction of current flow as shown in Fig. 2. 

The source and drain ferromagnets are magnetized along the channel (x-direction) 
and their magnetization vectors point in the same direction (parallel 
magnetization). The superlattice potential causes miniband formation for  
majority spins entering from the source and exiting at the drain. When the Fermi 
level is inside a miniband, the transmission through the device is relatively 
large and the transistor is ON. By changing $\delta (\Delta{E_c})$, one can move 
the miniband away from the Fermi level (or vice versa), thereby decreasing the 
transmission probability and turning the transistor OFF.

The optimal device will require minimal change $\delta (\Delta{E_c})$  to turn 
the transistor ON or OFF. Furthermore, in order to reduce fabrication 
complexity, we also prefer fewest repetitions of the superlattice unit (as few 
inter-digitated gates as possible). Here, we show that even a single period of 
the superlattice unit, implemented with just a single pair of gates, is 
sufficient to yield a very low switching voltage of about 2.6 mV, and yet a 
large conductance ON/OFF ratio of about 60, up to a temperature of 10 K. The 
analysis described here can be easily extended to the case of multiple periods. 
Use of multiple periods can enhance device performance by extending the 
temperature of operation to higher temperatures.

For simplicity, the conduction band diagram along the direction of current flow
is modeled as shown in Fig. 2. The contact potentials  are approximated by two 
positive delta-scatterers of strength $\Gamma$ at the contact/channel 
interfaces. This model is fairly accurate for heavy doping in the channel when 
the Schottky barriers at the contact/channel interfaces become very narrow.

The energy bands in the ferromagnets are modeled after the Stoner-Wohlfarth 
model which assumes that the majority and minority spin bands are split by an 
exchange energy $\Delta$. The Fermi level is  below the bottom of the minority 
spin band, so that the ferromagnets are 100\% spin polarized at low temperatures 
(i.e. they are half metallic). There is a Rashba spin orbit interaction in the 
channel because of the symmetry breaking electric field at the heterointerface, 
but we assume that it is independent of the gate potential since the latter is 
never varied by more than a few mV and the spin orbit interaction strength is 
weakly sensitive to gate potentials \cite{nitta}. 

We calculate the conductance through the device using the ballistic
model for spin transport developed in refs. \cite{cahay1,cahay2,cahay3}. The 
$y$- and $z$-components of the wavefunction in the channel
will be slightly different under the two gates for non-zero values of $\delta 
(\Delta E_c)$, but this subtlety is neglected here since $\delta (\Delta E_c)$ 
is never any more than just a few meV, as we show later. For the same reason, 
any difference between 
the Rashba-spin orbit coupling constants under the two gates is ignored.

\section{Results}

To calculate the linear response conductance of the SpinFET under various gate 
biases, we first calculate the spin dependent transmission probability of an 
electron through the device following the recipe of refs. 
\cite{cahay1,cahay2,cahay3}. This technique is fairly involved and the reader is 
referred to refs. \cite{cahay1,cahay2,cahay3} for the details. The results 
presented here are based on the assumption of  the device parameters listed in 
Table I.

Fig. 3 is a plot of the zero temperature conductance (in units
of $e^2/h$)  for the two following biasing
configurations: (a) the curve labeled ``1'' corresponds to the case
when the bias on the inner gate is fixed at
$\Delta{E_c}=\Delta{E_c}^{\star}$ = 4192 meV (which corresponds to a
transmission resonance when all gates are biased at the same potential), while 
the bias on the
outer gate is varied by 
$\pm \delta(\Delta{E_c})$ around the quiescent value of 4192 meV; (b) the curve 
labeled ``2'' corresponds to the second
biasing configuration when the bias on the outer
gate is kept fixed at $\Delta{E_c}=\Delta{E_c}^{\star}$ = 4192 meV
and the bias on the inner gate is varied by $\pm \delta(\Delta{E_c})$ around the 
quiescent value. 

A comparison of the curves labeled ``1'' and ``2'' reveals that it is
advantageous to operate the device with
$\delta(\Delta{E_c})$ = 2.58 meV, indicated by a vertical arrow in Fig. 3. 
With this value of $\delta(\Delta{E_c})$, switching from the biasing 
configuration ``1'' 
($\Delta{E_c}^{\star},\Delta{E_c}^{\star} - 2.58 meV,
\Delta{E_c}^{\star}$) to ``2'' ($\Delta{E_c}^{\star} - 2.58 meV,
\Delta{E_c}^{\star}, \Delta{E_c}^{\star} - 2.58 meV$) switches the device from 
ON ($G=e^2/h$) to OFF ($G\sim0$).

Fig. 4 is a plot of the  energy dependence of the transmission
coefficient $T(E)$  in  the two biasing configurations discussed above. 
Hereafter, we label these two configurations ``ON'' and ``OFF'' since they 
correspond to ON and OFF states of the transistor.
Both in the ON and OFF states, $T(E)$ contains several peaks
and troughs due to Ramsauer and Fano resonances in the conductance
of the channel, features which we have analyzed thoroughly in the past 
\cite{cahay1,cahay2,cahay3}.
In the energy range [4198.5 - 4200.5] meV, the transmission coefficient reaches 
unity 
several times when the SpinFET is biased in the ON configuration 
but is close to zero when biased in the complementary scheme.
Therefore, the energy range [4198.5 - 4200.5] meV constitutes the ``pass band'' 
in the ON configuration and also the ``stop band'' in the OFF configuration. 
Consequently, if we place the Fermi level within this energy range
(by appropriate channel doping, or even using a back gate), then we can switch 
the device ON or OFF by going from the one gate biasing scheme to the other. 
Since this requires changing the bias on any gate by at most 2.58 mV, the 
effective switching voltage $V_{switch}$ is only 2.58 mV.

We can estimate the dynamic energy dissipated during switching. Assuming that 
the gate capacitance $C$ (including interconnects) is about 1 fF, the maximum 
energy dissipated during a switching event is $(1/2)CV_{switch}^2$ = 3.3 
$\times$ 10$^{-21}$ Joules, which is 4-5 orders of magnitude smaller than what a 
typical transistor dissipates in the Pentium IV chip today \cite{sia}.

The full width at half maximum (FWHM) for $T(E)$ in the ON configuration 
(distance between
points $A^{'}$ and $B^{'}$ in Fig. 4) is roughly 1.5 meV. This is the effective 
width of the pass band. It 
 is about three times the distance between the points $A$ and $B$, which is the 
FWHM of
the transmission coefficient versus energy curve when the inner and outer gates 
are
biased at the {\it same} potential (corresponding to $\Delta{E_c} = 4192 meV$ 
under both gates). Therefore, the use of a dual gate, as opposed to a single 
gate, has broadened the transmission resonance by a factor of 3.

Fig. 5 shows  the transfer characteristic of the device (conductance versus gate 
voltage) at zero temperature. This plot displays 
the change in the conductance
as the potential on the inner gate is gradually increased from
 4192 meV to 4192 + $x$  meV while simultaneously reducing
the potential on the outer gate from 4192 + 2.58 meV 
to 4192 + 2.58 - $x$  meV.  We see that transistor has a fairly sharp switching 
characteristic -- the transition width is only about 2 mV -- which is the 
hallmark of a good inverter.  

The transistor performance obviously degrades at high temperatures because of 
thermal averaging over electron energy.
Thermal averaging can be viewed as a convolution of $T(E)$ with a function of 
width roughly
equal to 4 $k_BT $ \cite{bagwell}. Therefore, the device performance is not 
expected to deteriorate substantially
until $4 k_BT \approx 1.5 meV$, which is the distance between the  points 
$A^{'}$ and $B^{'}$ in Fig. 4. Setting $4kT$ = 1.5 meV, we find that the upper 
limit on the temperature is about 5 K. 
Over the energy range of 1.5 meV, the conductance in the OFF state stays very 
close to zero since $T(E)$
remains approximately zero in this energy range. Therefore, the dual-gate 
SpinFET will retain a large ON to OFF conductance ratio up to a temperature of 
at least 5 K.

The temperature dependence is shown in Fig. 6 where the conductance of the 
device is
plotted as a function of temperature for the two biasing
configurations discussed above. Table II lists the values of
the conductance ON/OFF ratio calculated as a function of temperature
for $\delta(\Delta{E_c})$ = 6 meV.
The ON/OFF ratio is about  60 at $T=10K$. This is comparable to what is 
typically obtained today with carbon nanotube transistors. 

One deleterious effect of high temperatures is that the maximum ON conductance 
has dropped to $0.27 e^2/h$. This can be remedied by using multiple parallel 
channels to increase the total device conductance, while still maintaining a 
large conductance ON/OFF ratio.

\section{Conclusion}

In conclusion, we have shown that using a dual gate configuration, it is 
possible to realize
short-channel SpinFETs with  large conductance ON/OFF ratio at temperatures 
above that of liquid helium.
These SpinFETs  possess excellent inverter characteristics, making 
them ideal for  low power logic circuits. The sharp turn-off behavior is 
conducive to good noise margin and restoration of logic levels at circuit nodes 
\cite{hodges} for fault-tolerant computing and signal processing.  The very 
small power dissipation will also allow extremely high integration density.

\newpage

\vskip .1in
\begin{center}
{\bf Table I: Parameters of spin interferometer}
\end{center}
\begin{center}
\begin{table}[h]
\centering
\begin{tabular}{cc} \hline\hline
Fermi Energy $E_F$ in contacts (eV) & 4.2         \\
Rashba spin-orbit coupling constant ${\alpha}_R$ ($10^{-11}$ eVcm)  &  1.    \\
Lande Factor $g^*$ &  -14.9   \\
Effective mass ${m_f}^* $ in Fe contact ($m_0$)  &   1.     \\
Effective mass ${m_s}^* $ in InAs channel ($m_0$) & 0.023\\
Total length of the channel(${\mu}m$) & 0.15\\
Strength of delta scatterer at the contact/channel interface (eV $\AA$) & 2.0\\
Exchange splitting energy $\Delta $ (eV) & 6.0\\
Magnetic field along the channel (Tesla)& 0.6\\
\hline\hline
\end{tabular}
\end{table}
\end{center}

\newpage
\vskip .1in
\begin{center}
{\bf Table II: Temperature dependence of conductance \\
of the SpinFET and ON/OFF ratio}
\end{center}
\begin{center}
\begin{table}[h]
\centering
\begin{tabular}{cccc} \hline\hline
Temperature (Kelvin)& $G_{ON} (e^2 /h)$ & $G_{OFF} (e^2 /h)$ & $G_{ON}/G_{OFF} 
$\\
1 & 0.72 & 0.007 & 102 \\
2 & 0.67 & 0.007 & 97\\
5 & 0.46 & 0.006 & 74\\
10& 0.27 & 0.005 & 60\\
\hline\hline
\end{tabular}
\end{table}
\end{center}

\newpage
\parindent 0cm
\vskip .1in

{\bf Figure 1}: (Top) Cross-sectional view of the SpinFET composed
of two interdigitated gates controlling the current flowing between
two half metallic contacts. (Bottom) Top view of the dual-gate SpinFET 
consisting of an array
of parallel quasi one-dimensional channels (dotted lines).
The magnetization in the contacts is assumed to be in the direction ($+x$) of 
current flow.

\vskip .1in
{\bf Figure 2}: Energy band diagram along the direction of one of the channels 
of the SpinFET.
In the half-metallic contacts, the exchange energy $\Delta$ is large enough that 
the bottom of the minority spin band is above the Fermi level, so that the spin 
polarization in the ferromagnet is 100\%. The minority spins are evanescent in 
the channel. The energy bands in the ferromagnets are modeled using the 
Stoner-Wohlfarth model.  
The barrier height $\Delta E_c$ under each gate is controlled by the applied 
bias. In the calculations, 
$\Delta E_c$ is assumed to include the effects of the quantum confinement in the 
y- and z-directions.
The contact potentials at the ferromagnet/semiconductor interface are modeled as 
simple one-dimensional
delta-potentials. $\Delta E_c$ is set equal to $ \Delta{E_c}^{\star} $ = 4.192 
eV under one
of the gates and varied by $ \delta ( \Delta E_c )$ in either direction under 
the other gate.

\vskip .1in
{\bf Figure 3}: Zero temperature conductance  as a function 
of $ \delta (\Delta E_c )$ for the two different biasing configurations 
discussed in the
text. The parameters of the SpinFET are listed in Table I.

\vskip .1in
{\bf Figure 4}: Transmission coefficient versus energy of majority spins in the 
ON and OFF biasing configurations. The dot-dash curve corresponds
to the case where the same potential is applied to both gates
such that $\Delta E_c$ =  $ \Delta{E_c}^{\star} $ = 4.192 eV.

\vskip .1in
{\bf Figure 5}: Zero temperature inverter characteristic of the SpinFET. The 
conductance is plotted as a function of the incremental bias $x$ on either gate 
assuming a quiescent gate bias of 4.192 V.

\vskip .1in
{\bf Figure 6}: Temperature dependence of the conductance 
versus $ \delta (\Delta{E_c}) $  in the ON and OFF states.
The parameters of the SpinFET are listed in Table I and $\Delta E_c$ 
=  $ \Delta{E_c}^{\star} $ = 4.192 eV.

\newpage
\
\vskip .2in
\begin{figure}[h]
\centerline{\psfig{figure=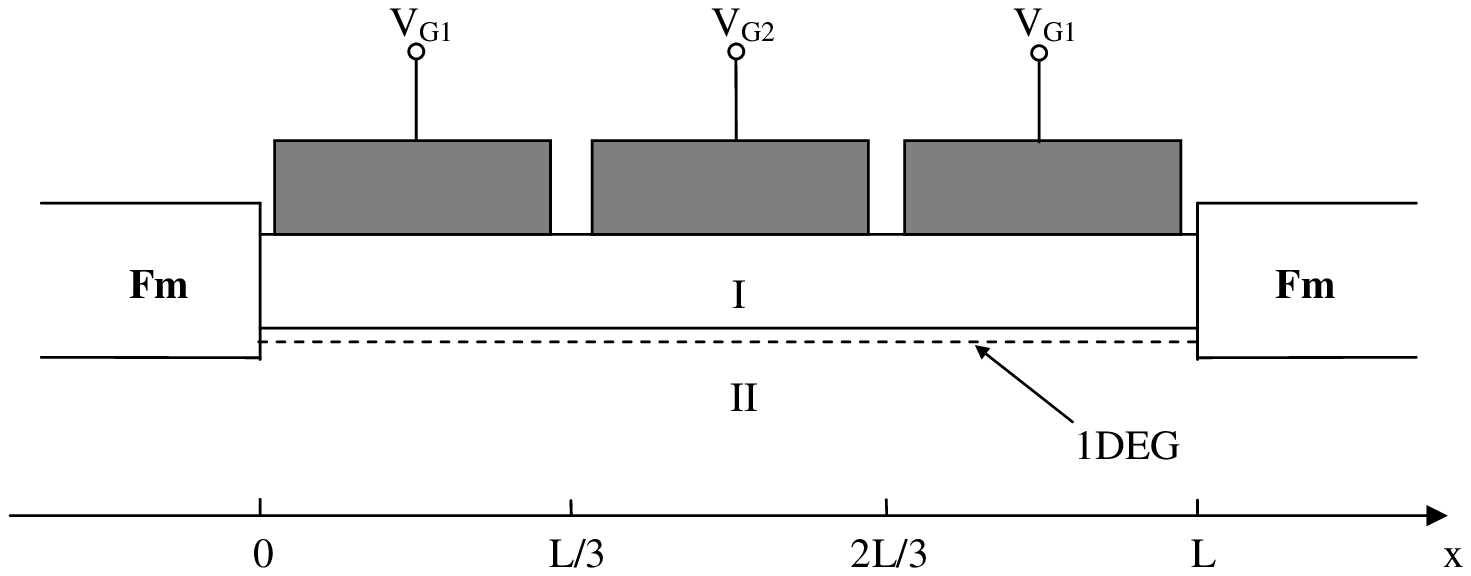,height=3in,width=5.5in}}
\end{figure}

\
\vskip .1in
\begin{figure}[h]
\centerline{\psfig{figure=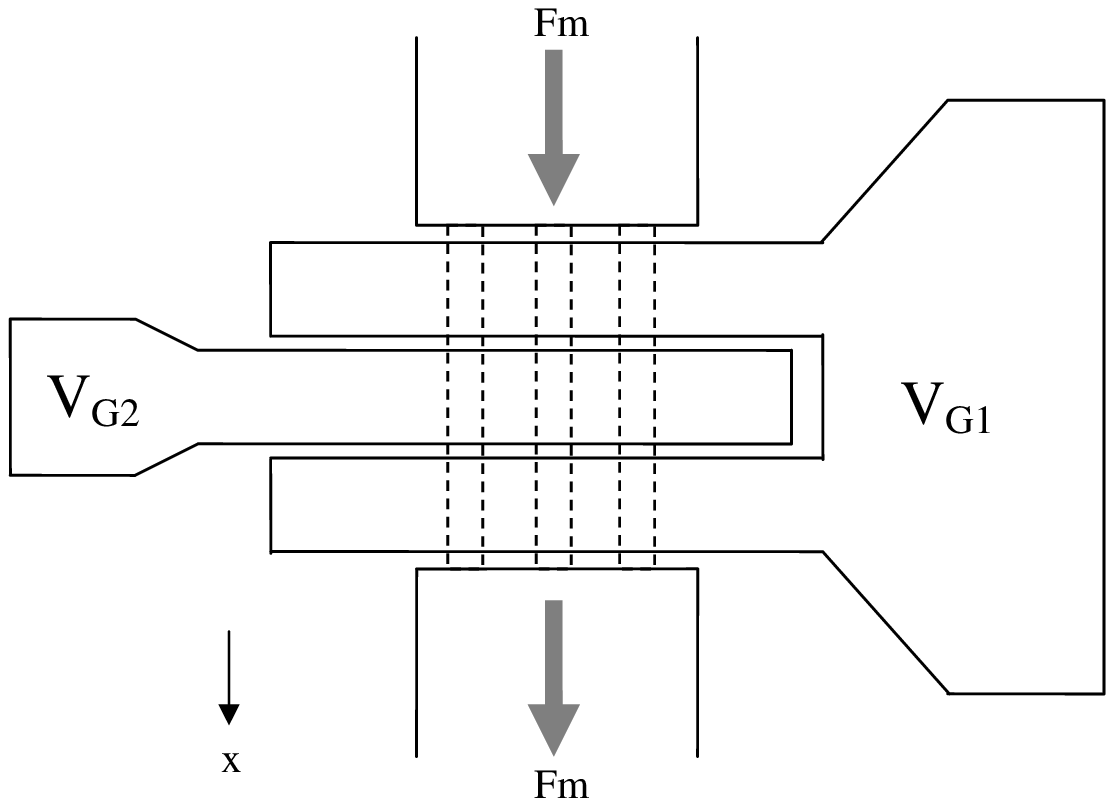,height=3in,width=5.5in}}
\end{figure}
\begin{center}
\vskip .3in
{\bf Figure 1}
\end{center}

\newpage
\
\vskip .2in
\begin{figure}[h]
\centerline{\psfig{figure=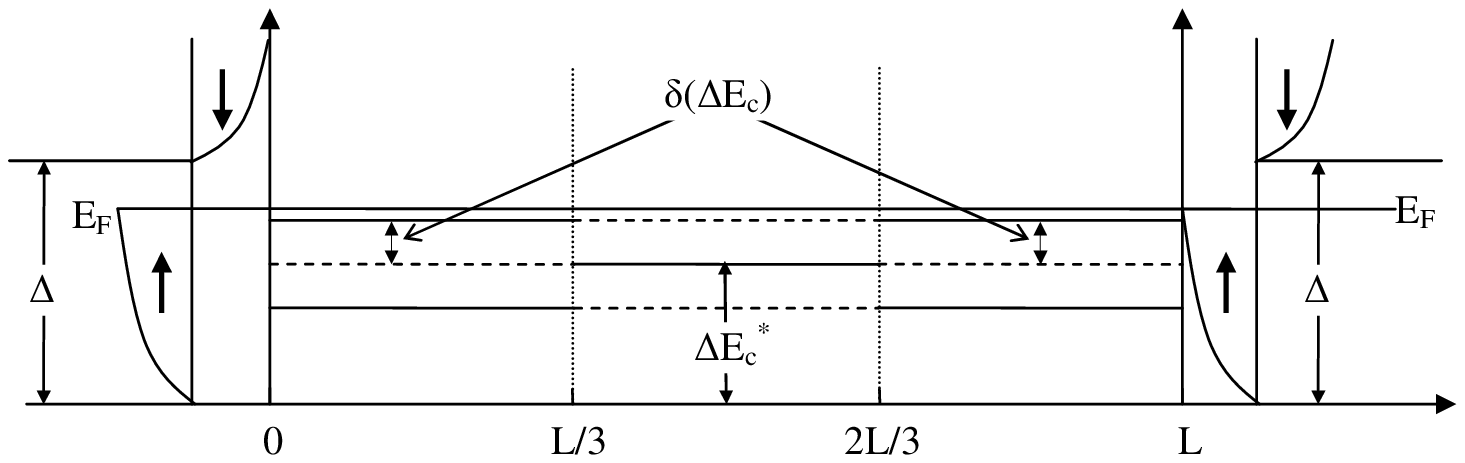,height=3in,width=5.5in}}
\end{figure}
\begin{center}
\vskip .3in
{\bf Figure 2}
\end{center}

\newpage
\vskip 0.2in
\begin{figure}[h]
\centerline{\psfig{figure=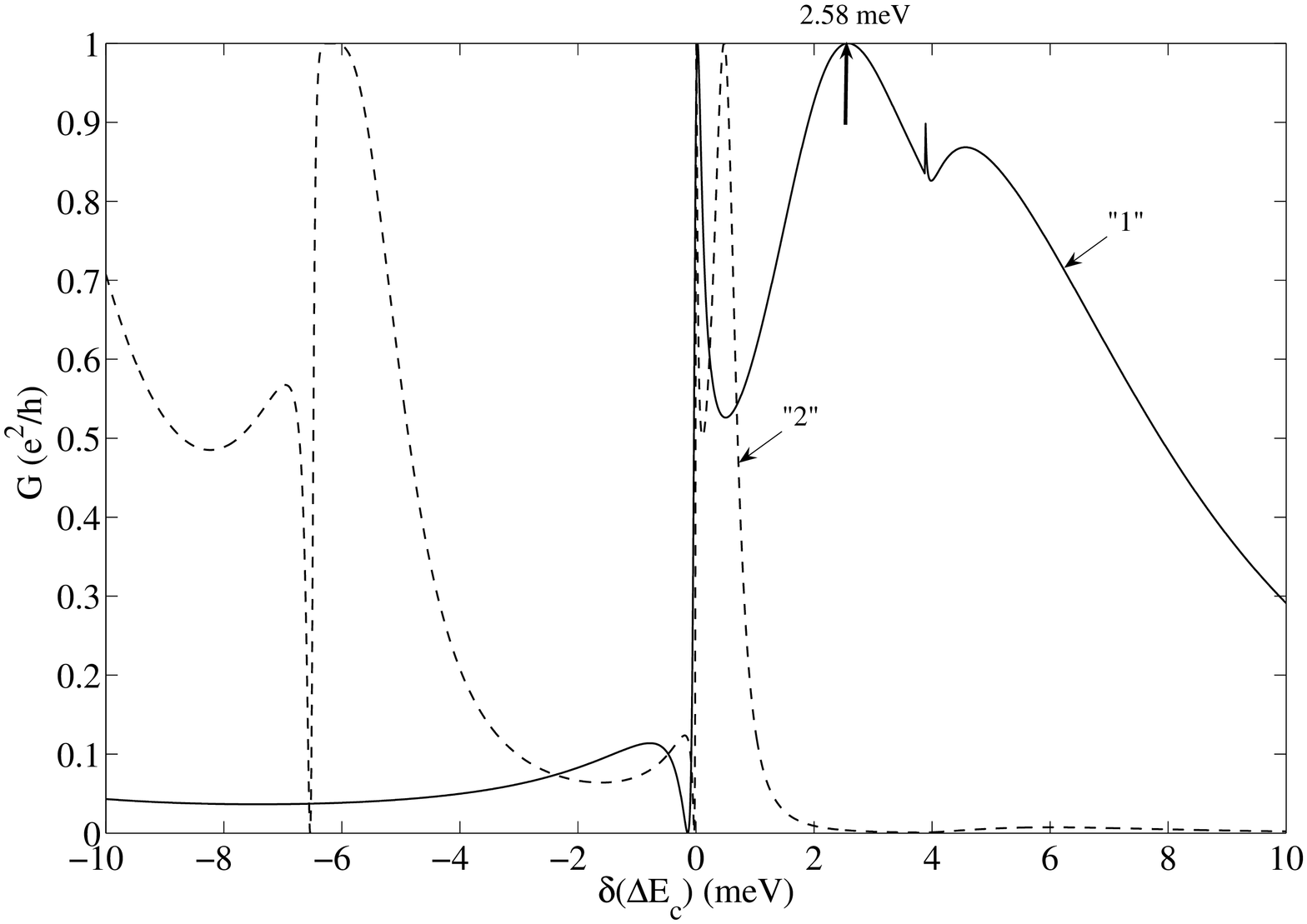,height=4.5in,width=4.5in,angle=-90}}
\end{figure}
\vskip .1in
\begin{center}
{\bf Figure 3}
\end{center}

\newpage
\
\vskip 0.2in
\begin{figure}[h]
\centerline{\psfig{figure=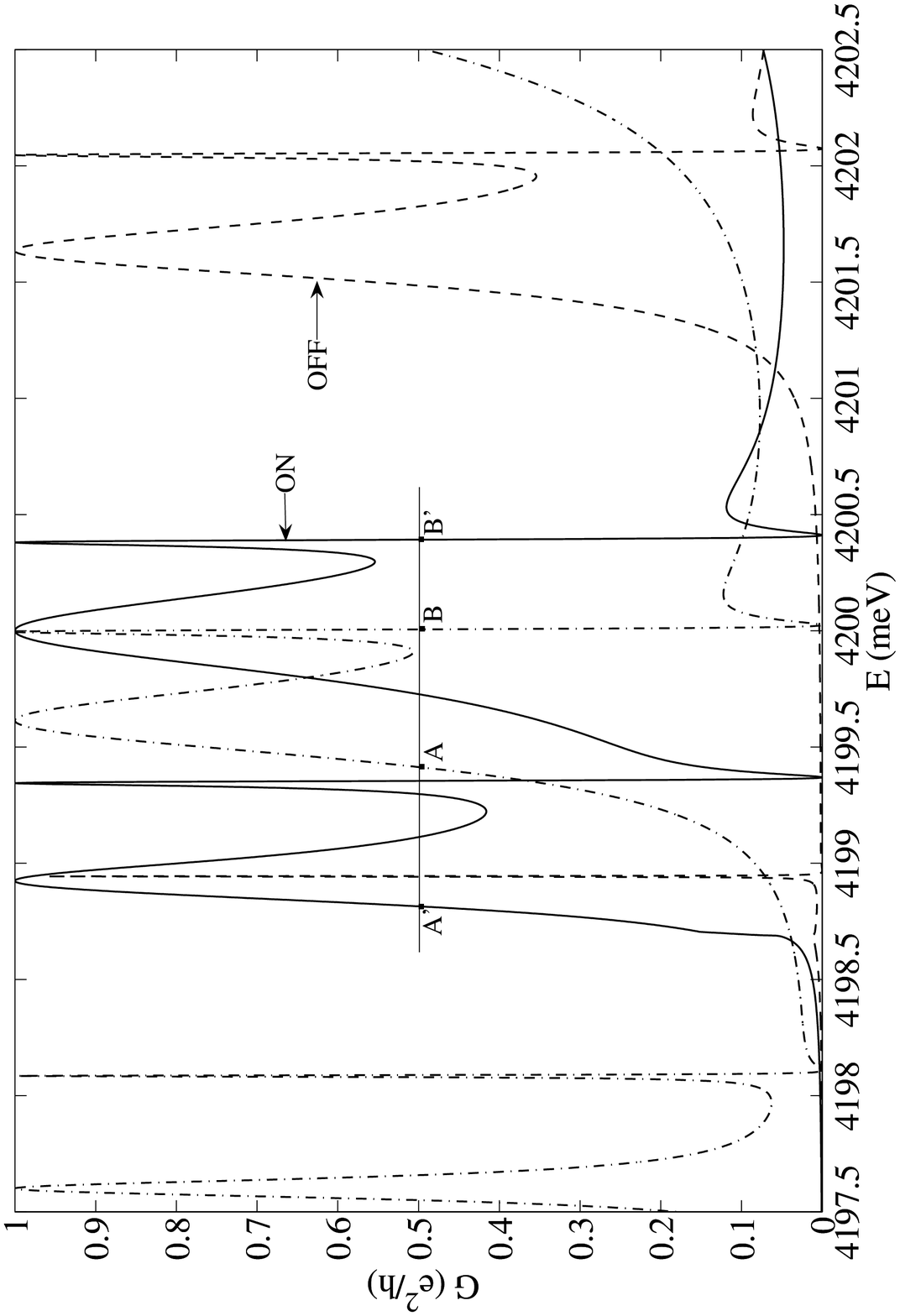,height=6.1in,width=6in,angle=-90}}
\end{figure}
\vskip .2in
\begin{center}
{\bf Figure 4}
\end{center}

\newpage

\vskip .1in
\begin{figure}[h]
\centerline{\psfig{figure=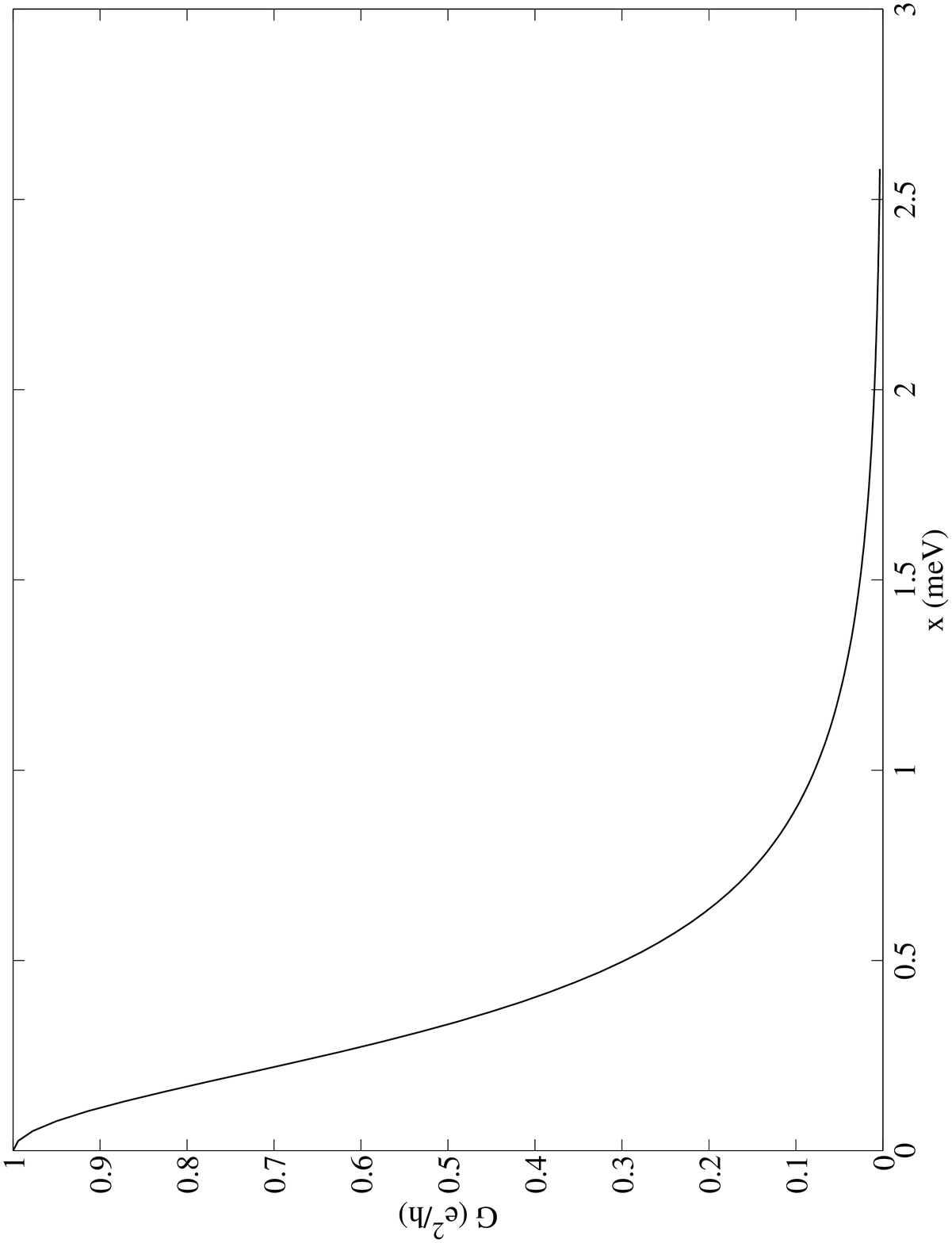,height=4.5in,width=4.5in,angle=-90}}
\end{figure}
\vskip .2in
\begin{center}
{\bf Figure 5}
\end{center}

\newpage
\vskip 0.2in
\begin{figure}[h]
\centerline{\psfig{figure=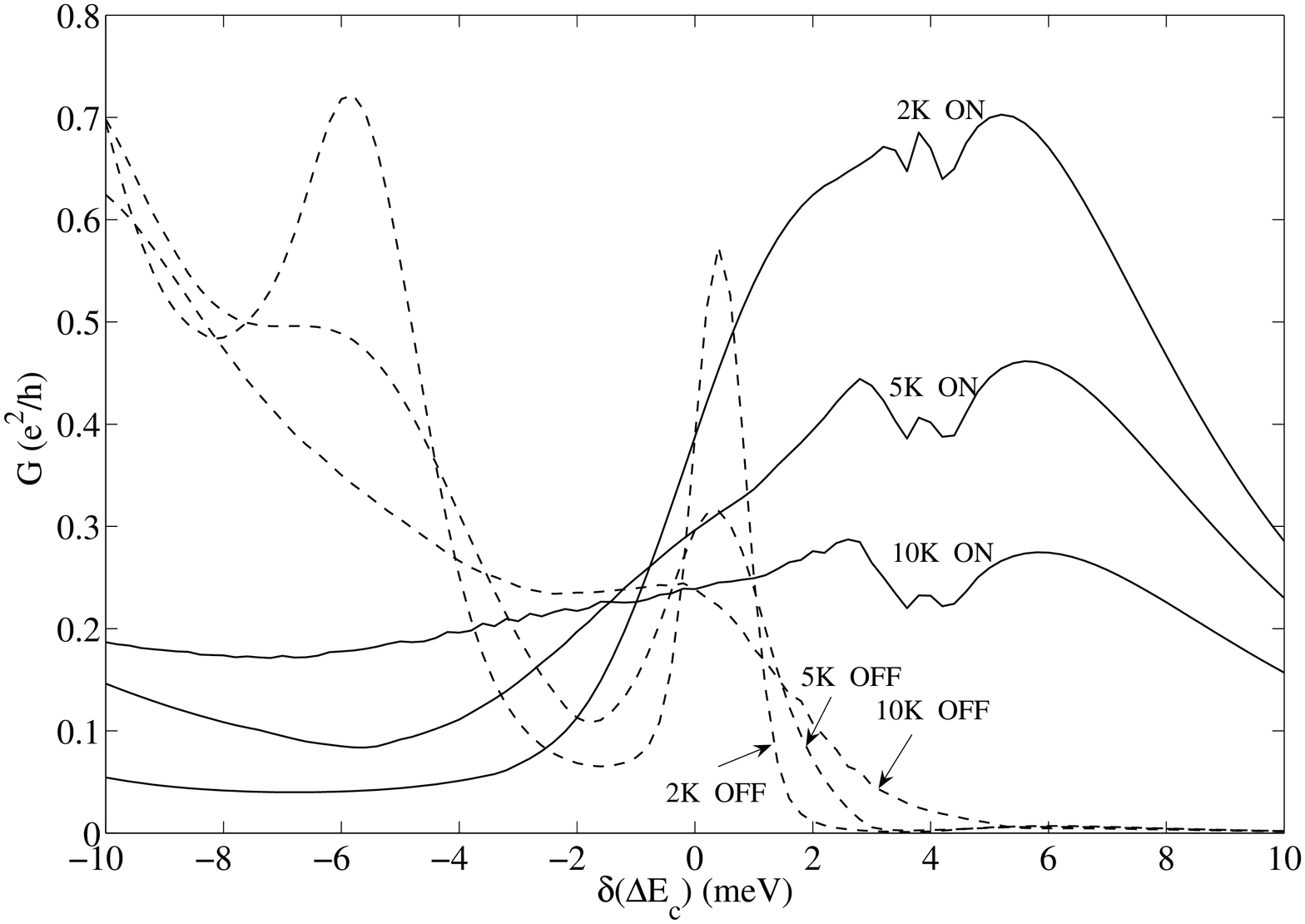,height=5.5in,width=6in,angle=-90}}
\end{figure}
\vskip .2in
\begin{center}
{\bf Figure 6}
\end{center}

\end{document}